\documentclass[reprint,amsmath,amssymb,aps]{revtex4-1}

\usepackage{graphicx} 
\usepackage{dcolumn}
\usepackage{bm}
\usepackage{natbib}
\usepackage{amsmath}
\usepackage{tcolorbox}

\begin{document}

\title{Multi-mode four-wave mixing with a spatially-structured pump}

\author{Jon D. Swaim}
\author{Erin M. Knutson}
\author{Onur Danaci}
\author{Ryan T. Glasser}
\email{rglasser@tulane.edu}
\affiliation{Physics and Engineering Physics Department, Tulane University, New Orleans, LA 70118, USA}

\date{\today}

\begin{abstract}
We demonstrate a new four-wave mixing (4WM) geometry based on structured light.  By utilizing near-field diffraction through a narrow slit, the pump beam is asymmetrically structured to modify the phase matching condition, generating multi-mode output in both the spatial and frequency domains.  We show that the frequency parameter enables selection of various spatial-mode outputs, including a twin-beam geometry which preserves relative intensity squeezing shared between the two beams.  The results suggest that the engineering of atomic states via structured light may provide a pathway to a diverse set of quantum resources based on multi-mode squeezed light.
\end{abstract}

\maketitle

\clearpage

Spatially-structured light is emerging as a promising new tool for manipulating nonlinear and quantum optical phenomena in atomic media~\cite{Arlt_1999, Barreiro2004, Andersen2006, Walker2012, Fickler2012, Akulshin2015, Zhang2016, Danaci2016, Wang2017, Jia2017, Rubinsztein-Dunlop2017, Lanning2017}.  Some notable examples of structured (coherent) light have included optical vortices~\cite{Nye1974, Soskin1997}, orbital angular momentum states~\cite{Allen1992, Mair2001, Molina-terriza2007}, tailored phase and intensity patterns and dynamical amplitude and phase control of light~\cite{Rubinsztein-Dunlop2017}, all of which have been facilitated by devices such as the spatial light modulator.  These developments have since led to enhanced forms of microscopy~\cite{Sheppard1997, Mosk_2012, Hell2015, Taylor2015} as well as metrology~\cite{Treps2002, Belmonte2011}.  In the context of quantum resources, multi-mode quantum imaging and entanglement require quantum light which is tailored in the spatial degree of freedom, and as such lend themselves to nonlinear techniques which are inherently multi-mode (i.e., without cavities).  One such technique is 4WM in hot atomic vapors~\cite{McCormick2007, Guo2014, Swaim2017}.  There, a strong $\chi^{(3)}$ nonlinearity generates multiple quantum-correlated beams which can be shaped spatially via structuring of the incident pump light~\cite{Wang2017, Jia2017}.  Recently, with the use of two tilted coherent pumps, this approach enabled measurement of relative intensity squeezing shared among combinations of six spatially-separated beams~\cite{Wang2017}.  These advancements hold promise for being able to control the nature of multi-mode quantum correlations at the source, without introducing additional losses or alignment issues.


\begin{figure}[b!]
\begin{center}
\includegraphics*[width=\columnwidth]{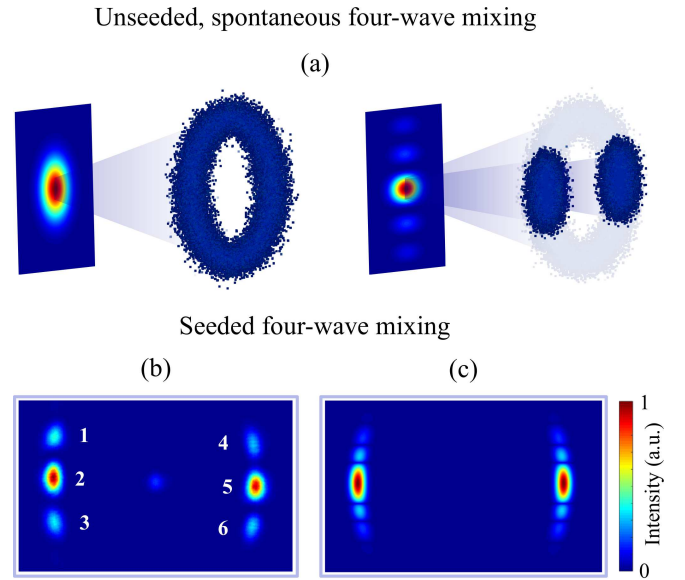}
\caption{\label{concept} Multi-mode four-wave mixing with structured light.  (a) Unseeded, spontaneous four-wave mixing geometries for a circularly symmetric pump (left) and an  asymmetrically structured pump (right).  (b)  Experimental image ($4.4$ mm $\times$ $2.75$ mm) of four-wave mixing when the process is seeded with a weak input probe. (c) Theoretical far-field intensity pattern for our proposed soft gain-medium aperture.  }
\end{center}
\end{figure} 

In this Letter we expand upon these concepts, and show that a simple diffractive element (a slit) can be combined with a nonlinear optical device (4WM in atomic vapor) to create light which is structured in multiple degrees of freedom, using only a single pump beam.  In a typical 4WM scenario, a hot vapor of alkali atoms is strongly pumped with a circularly symmetric beam, leading to the spontaneous emission of a ring of phase matched 4WM photons (referred to as probe and conjugate), as shown in the left of Fig.~\ref{concept}a.  In the present experiment, we instead pump the atoms with light which is diffracting from passing through a nearby rectangular slit, resulting in a restricted phase matching condition due to the broken circular symmetry and strong spatial dependence in the propagation direction (right of Fig.~\ref{concept}a).  When the experiment is seeded with a weak, Gaussian-shaped input probe, the spatial phase mismatch between the structured pump and the input probe then leads to multi-mode output, both in space (as shown in Fig.~\ref{concept}b) as well as in the frequency domain.  In general, the spatial-mode outputs are highly symmetric and tend toward a six-fold geometry, though a variable number of spots can be achieved depending on the experimental parameters.  We model these effects numerically, and find that the results can be understood by considering the structured pump as an effective gain aperture.  As shown in Fig.~\ref{concept}c, the numerical predictions qualitatively match those which are observed in the experiment.    Additionally, rather than observing individual probe and conjugate resonances when scanning the two-photon detuning $\delta$, we instead find a doublet of modes for each of the six spots, with a tunable splitting frequency on the order of the individual linewidths ($\sim 10$ MHz).  Lastly, we show that the observed frequency modes correspond to different spatial-mode outputs with unique intensity correlations, one of which is a twin-beam geometry that preserves relative intensity squeezing between the two beams.

\begin{figure}[t!]
\begin{center}
\includegraphics*[width=\columnwidth]{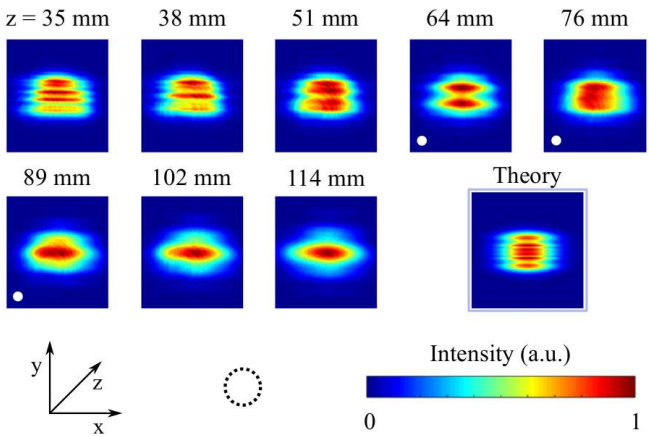}
\caption{\label{pump} Spatial profiles of the diffracting pump.  Experimental intensity images ($1.1$ mm $\times$ $1.1$ mm) of the pump along the propagation coordinate $z$, relative to the position of the slit.  The profiles at $z = 64$ mm, $76$ mm and $89$ mm coincide with the length of the vapor cell, and the pump-probe crossing occurs near $z \sim 76$ mm.  The theoretical (near-field) pump profile used in the soft aperture model is also shown, and the relative size of the probe is shown with a dotted black circle.}
\end{center}
\end{figure} 

In the experiment, coherent light derived from a Ti:Sapph laser is collected into a single-mode fiber to produce a clean, approximately Gaussian-shaped pump beam ($1/e^2 \sim 900$ $\mu$m).  The input probe is generated by double-passing a portion of the light through an acousto-optic modulator operating around $1.52$ GHz (roughly half the ground state hyperfine splitting of $^{85}$Rb), and shaping it into a Gaussian profile $400$ $\mu$m in diameter.  The power of the input probe is $\sim 100$ $\mu$W.  As in previous experiments on 4WM in the double-$\Lambda$ configuration~\cite{McCormick2007, Guo2014, Swaim2017}, the pump and probe are made to overlap in the center of a $25$ mm long atomic vapor cell at a small angle ($0.6 ^\circ$), leading to an amplified probe and generated conjugate.  To structure the pump, it is passed through a rectangular slit ($d \sim 530$ $\mu$m) positioned $76$ mm before the center of the cell.  Strong 4WM is then achieved for a pump power of $200$ mW and cell temperature of $T \sim 125$ $^\circ$C, with the pump detuned to the blue side of the $^{85}$Rb D1 line by $\Delta \approx 2$ GHz.  After filtering out the pump using a polarizing beam splitter, the probe and conjugate beams are imaged with a camera placed $140$ mm after the center of the vapor cell, and detected on a balanced photodetector (using an RF gain of 10$^5$ V/A).  

\begin{figure}[b!]
\begin{center}
\includegraphics*[width=\columnwidth]{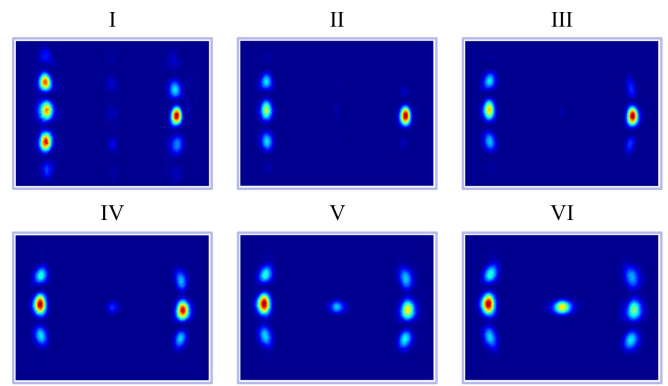}
\caption{\label{outputs}Spatial-mode outputs for various pump detunings.  The normalized intensity images labeled I-VI are taken for pump detunings of $\Delta = 0.8$, $1.0$, $1.3$, $1.5$, $1.8$ and $2.0$ GHz, and $\delta = - 5$ MHz.  The intensity scale is the same as in Fig.~\ref{pump}. }
\end{center}
\end{figure} 

\begin{figure}[t!]
\begin{center}
\includegraphics*[width=\columnwidth]{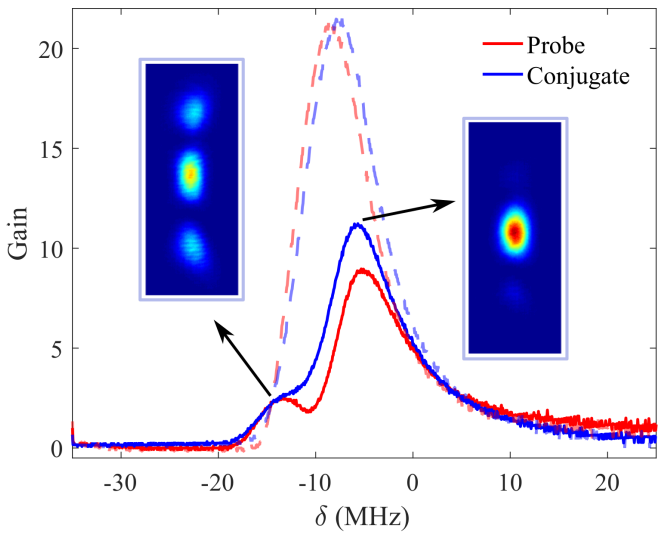}
\caption{\label{resonances} Four-wave mixing resonances.  The asymmetrical structuring of the pump results in a doublet of modes for each probe and conjugate (solid lines).  For comparison, the gain lines obtained with a circularly symmetric pump are shown with dashed lines.  The spatial structures of the conjugates at resonance are shown in the insets.  The spectra are measured over probe and conjugate spots $I_2$ and $I_5$ only. }
\end{center}
\end{figure}

We find that near-field diffraction provides an efficient means of symmetrically imprinting spatial information on to the probe and conjugate beams.  In Fig.~\ref{pump}, we show intensity profiles of the diffracting pump along the propagation coordinate $z$, with white circles indicating the profiles which coincide with the vapor cell.  Building on modeling from our previous work~\cite{Danaci2016}, we calculate the expected output 4WM mode profiles from this type of setup, where the interaction is modeled as a soft aperture dependent upon the spatial profile of the pump and constrained by the phase matching conditions.  To arrive at the result shown in Fig.~\ref{concept}c, we used as the aperture the near-field diffraction pattern of a Gaussian beam passed through a rectangular aperture (i.e., the theoretical pump profile shown in Fig.~\ref{pump}).  We see that the model predicts qualitatively similar output patterns as those observed in the experiment, and we expect that even better agreement could be obtained by taking into account the dynamical interplay between diffraction and gain.  Ultimately, the spatial phase mismatch responsible for multi-mode output is due to the overlap and crosstalk between the pump and probe beams, as well as the frequency-dependent refractive index of the medium.  Shown in Fig.~\ref{outputs}, various intensity output patterns can then be achieved depending on the one-photon detuning.  In particular, we see that further away from resonance the number of resolvable spots tends to decrease, and the changing refractive index also introduces some small curvature in the output.  Both of these effects can be viewed as being a result of the spatially-modified phase matching condition.

\begin{figure}[b!]
\begin{center}
\includegraphics*[width=\columnwidth]{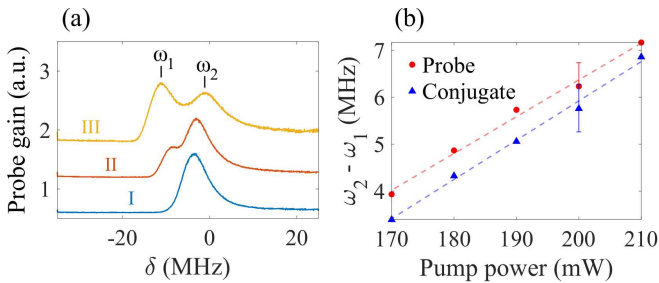}
\caption{\label{splitting} Tunable frequency splitting.  (a) The probe resonances, labeled I-III, are shown for slit sizes of $d \sim 580$ $\mu$m, $530$ $\mu$m and $455$ $\mu$m, respectively. (b)  The frequency splitting between the two modes exhibits a linear dependence on the optical power of the pump.  The data in (b) is taken with $d \sim 530$ $\mu$m.}
\end{center}
\end{figure} 

Another consequence of the phase mismatch is that each bright spot in the output consists of multiple 4WM resonances in frequency space.  In Fig.~\ref{resonances}, we show the measured gain resonances for the probe and conjugate spots labeled $5$ and $2$ in Fig.~\ref{concept}b, respectively, for our structured pump (solid lines), as well as for the typical case of a circularly symmetric pump (dashed lines).  The act of introducing spatial structure into the pump splits a single 4WM resonance into a frequency doublet, with a splitting of about $8.5$ MHz in the present case.  As pointed out in Ref.~\cite{Zhu2017}, there exists a direct correspondence between this splitting (i.e., multi-mode output) and the spatial phase mismatch, which experimentally we find can be achieved by obstructing the pump in a number of ways.  Also, it is worthwhile to mention that the splittings observed in Refs.~\cite{Zhang2015, Zhu2017} were obtained using a different method, and did not involved a structured pump beam.  Although the amount of splitting differs among the various spots, these resonances in general correspond to different spatial-mode outputs.  To illustrate this, in the insets of Fig.~\ref{resonances} we show images of the conjugates' spatial profiles measured at the two resonance frequencies.  When we detune the probe to the lower frequency resonance of the doublet ($\delta \sim -14$ MHz), we obtain the six-spot geometry.  However, when it is tuned to the higher frequency one, we obtain the more familiar case which is expected for an unobstructed, Gaussian pump.  To our knowledge, this is the first demonstration of the ability to switch between multi-mode outputs in such a manner.

\begin{figure}[t!]
\begin{center}
\includegraphics*[width=\columnwidth]{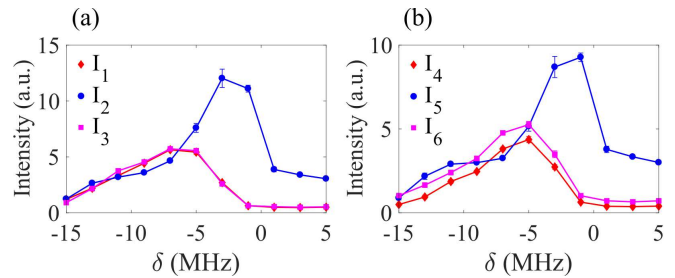}
\caption{\label{composition} Intensity composition of the spatial-mode output vs. two-photon detuning.  Integrated camera intensities for (a) conjugates $I_1$, $I_2$ and $I_3$ and (b) probes $I_4$, $I_5$ and $I_6$.  For each data point, the intensities have been averaged from ten images, using an area of $\sim 1.4$ mm$^2$ for integration.  Intensity correlations exist between $I_2$ and $I_5$, as well as between $I_1$, $I_3$, $I_4$ and $I_6$.}
\end{center}
\end{figure} 

In such an experiment, it is important to be able to resolve the different frequency modes.  As a rule of thumb, this occurs when the splitting frequency is on the order of the individual linewidths of the resonances.  Now we show that the splitting frequency can be varied with the properties of the incident pump beam.  In Fig.~\ref{splitting}a,  we show the measured spectra for the probe (spot $5$) for three different slit sizes, with a fixed pump power of $200$ mW.  As the slit size is decreased, the elliptical anti-nodes of the pump spread out vertically, exaggerating the phase mismatch and producing a larger frequency splitting.  We also see that the relative strengths of the two modes (denoted $\omega_1$ and $\omega_2$) can be varied.  This property is useful because it allows for the two frequency modes to be easily resolved, in such a way that the gains are roughly equal (curve III in Fig.~\ref{splitting}).  On the whole, though, the 4WM gain decreases with smaller slit size.  Alternatively, the frequency splitting can be increased via light shifts induced by the pump beam.  In Fig.~\ref{splitting}b, we show the measured splitting $\omega_2$ - $\omega_1$ for the probe and conjugate spots $5$ and $2$, as a function of the pump power using a fixed slit size.  As expected, both exhibit a linear dependence~\cite{Levi2016}.  

Until now, we have only considered the modal composition of individual probes and conjugates in the output.  In search of spatial intensity correlations, we next analyze the composition of entire images as a function of the two-photon detuning.  In Fig.~\ref{composition}, we show the relative intensities for each spot obtained in the six-fold geometry shown in Fig.~\ref{concept}b.  While the central probe and conjugate ($I_2$ and $I_5$; blue circles) clearly track each other, the remaining four spots are also uniquely correlated.  Additionally, we find that the central spots are associated with both $\omega_1$ and $\omega_2$, whereas the other spots arise from the resonance at $\omega_1$.   Thus, in addition to switching between various spatial outputs (as demonstrated in Fig.~\ref{outputs}), the frequency parameter can be used to discriminate between various \emph{modes}, each with a unique intensity-correlation signature. 

\begin{figure}[t!]
\begin{center}
\includegraphics*[width=\columnwidth]{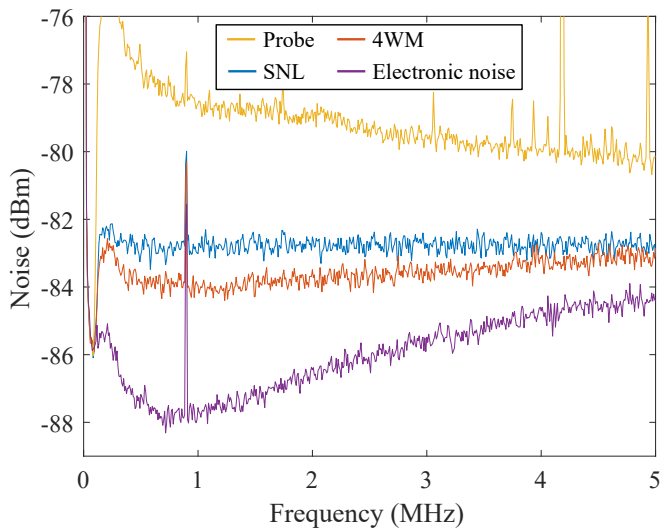}
\caption{\label{squeezing} Quantum correlations shared between twin beams in the high-frequency mode.  Spectrum analyzer traces, in ascending order, for electronic noise, the 4WM difference signal $I_2$ - $I_5$ (with $45$ $\mu$W of probe and $25$ $\mu$W of conjugate), the equivalent SNL and the probe noise only. Squeezing is observed for sideband frequencies from $0.1 - 5$ MHz, with $\delta = 13$ MHz.  The resolution and video bandwidths are $10$ kHz and $300$ Hz, respectively.  }
\end{center}
\end{figure} 

To further demonstrate the separability of the modes, we now show that the high-frequency mode preserves relative intensity squeezing between the probe and conjugate.  To do this, we compare the noise in the difference signal from the balanced photodetector ($I_2$-$I_5$) with the shot noise limit (SNL) for an equivalent amount of total optical power.  As shown in Fig.~\ref{squeezing}, the 4WM difference signal (red trace) drops below the SNL (blue) for sideband frequencies ranging from $0.1 - 5$ MHz, with a maximum squeezing of $1.2$ dB around $1$ MHz.  For completeness, we verified that the probe alone exhibits excess noise (yellow trace) and resembles a thermal state, as the quantum correlations are intensity-relative in nature.  Lastly, as a final check, we confirmed that the measured optical noise is above the electronic noise. 

To summarize, we have demonstrated a method of generating multi-mode output based on transferring spatial information from a single pump beam into spatial and frequency structuring of twin-correlated optical beams.  These results are encouraging, since the ability to increase the number of spatio-temporal degrees of freedom could ultimately lead to additional quantum modes~\cite{Devaux2000} -- a desirable consequence from the point of view of quantum imaging and communications~\cite{Devaux2000, Boyer2008}.  While in this work we focus on a geometry which splits the 4WM resonance into a frequency doublet, it is rather straightforward to modify the pump in order to achieve more than two resonances.  This could be achieved, for example, by replacing the slit with a spatial light modulator or a similar device.  In addition to finding potential quantum applications, these results also expand on the outlook for structured light.  In a recent review~\cite{Rubinsztein-Dunlop2017}, the authors comment: ``Going forward one wonders what exciting prospects await the expansion of structured light concepts beyond the spatial domain, for example, the shaping of light's time envelop and frequency control. Could we see optical fields structured in all dimensions and in all degrees of freedom?"

\bigskip

\textbf{Funding.} This research was supported by the Louisiana State Board of Regents and Northrop Grumman $\emph{NG - NEXT}$.  E. M. K. acknowledges funding from a National Science Foundation Graduate Research Fellowship.


%


\end{document}